# Framework for Formal Modelling of Metaverse Applications Using Hierarchical Colored Petri Nets

**Maryam Amin[1], Umara Noor[1], Zahid Rashid[2], Jörn Altmann[2, 3, 4]**

[1] Department of Software Engineering, Faculty of Computing and Information Technology,
International Islamic University, Islamabad, Pakistan
[2] Technology Management Economics and Policy Program, College of Engineering,
Seoul National University, Seoul, South Korea
[3] Institute of Engineering Research (IOER), College of Engineering,
Seoul National University, Seoul, South Korea
[4] Integrated Major in Smart City Global Convergence, Seoul National University, Seoul, South Korea

010481phdses25@student.iiu.edu.pk, umara.zahid@iiu.edu.pk, rashidzahid@snu.ac.kr, jorn.altmann@acm.org

**Abstract**

The Metaverse emerges by integrating highly-distributed, complex, and interconnecting technologies. These technologies need to be formally verified and evaluated through formal modelling before executing them in real-world applications, in order to avoid negative impacts on the real world due to failure of the Metaverse technologies. However, the formal modelling of Metaverse technologies is challenging due to its highly complex nature. Therefore, a comprehensive formal verification of the Metaverse technologies is needed for its realization in multiple potential areas. In this study, a framework is proposed for the formal modelling of Metaverse technologies, which allows holistic insights for all applications of Metaverse technologies. By utilizing the proposed framework, Metaverse applications of any complexity can be modeled. The working of the proposed framework is illustrated by modelling a case study of an Air Traffic Control system. In the proposed framework, we utilize hierarchical colored Petri nets for formal modelling of behavior of the air traffic control system. The correctness of air traffic control system properties, such as liveness, reachability, and boundedness, is verified in the proposed framework. The results of the case study reveal that the proposed framework can be used as a template for mathematical verification of challenging and complex Metaverse applications. The results also show that formal modelling provides an effective tool for identifying flaws in the early phases of the design of Metaverse applications. The implication of using formal verification is that it can increase confidence about the correctness of the Metaverse applications.

## 1. Introduction

Metaverse is an adaptation of the Internet, constituting seamless integration of interoperable, immersive, and un-divided (shardless) virtual 3D environments. In the Metaverse, physical world users are controlling avatars of virtual worlds. Damar et al. [43] describe the Metaverse as: "the layer between you and reality", and the Metaverse is referred to as a "3D virtual shared world where all activities can be carried out with the help of augmented and virtual reality services". The Metaverse is expected to have a huge impact on many aspects of how we live, work and socialize [12]. The Metaverse concept as outlined by Mark Zuckerberg describes "an integrated immersive ecosystem, in which the barriers between the virtual and real worlds are seamless to users, allowing the use of avatars and holograms to work, interact and socialize via simulated shared experiences" [64].

Recently, there has been a rapid increase in recognition of the Metaverse by the research community in areas like gaming, education, tourism, real estate, entertainment, office work, social life, healthcare, real estate, marketing, hospitality, and citizen-government interaction [64]. The adaption of the Metaverse is expected to have a transformational impact on business models, manufacturing, logistics processes, and operations management [64, 29, 14]. Metaverse based operations management combines both physical and digital aspects in manufacturing, supply chains, and logistics processes [14, 17, 18, 19, 29, 55]. The relevant stakeholders will have the leisure to perform informed decision-making based on the evolved operations management [17, 18].

Giant tech companies like Microsoft, Amazon, and Meta are working on Metaverse implementation in multiple real-world applications [35]. Other organizations also started to assess the potential for integrating Metaverse related technologies into operations management and business models.

However, formal modelling of Metaverse applications is limited, and there is no generic framework available for formal modelling of Metaverse applications. Existing studies focus on Metaverse architecture and its applications in specific domains but are limited in its modelling and verification. For example, prior research provided a Metaverse framework only pertaining to education [10, 26], marketing [41, 52] and advertisement [31]. Other studies [21, 28, 47, 64] proposed Metaverse architectures focusing on a specific point of interest, such as technology, security, and immersion. The literature survey shows that a comprehensive Metaverse framework, which can be applied to multiple domains is lacking. In addition, at present, there is a need for associated discussions about formal modelling and verification of Metaverse applications.

The objectives of this research are (i) to integrate the potential strengths of formal modelling and verification into the emerging field of the Metaverse, (ii) to develop a modelling framework to allow a holistic insight into Metaverse applications development and (iii) to demonstrate the practical applicability of proposed framework through a complex and novel case study.

In this research work, we extended the Metaverse architecture presented in [47] by adding necessary components of Metaverse-related technologies. For the modelling framework, which is part of the Metaverse architecture, we developed a formal model using hierarchical coloured Petri nets (CPN) Metaverse. Petri nets are an important tool for modelling and analysing the dynamic behaviour of complex systems [32]. In literature, Petri nets are applied for the formal modelling of complex systems like manufacturing system, communication protocols, and AI systems [58]. To ensure the model's correctness, CPN tools are used to verify the ATC Metaverse by analysing its state space graph and evaluating key properties such as reachability, liveness, and boundedness.

The properties of liveness, reachability, and boundedness of the Metaverse modelling framework are mathematically defined for this study and verified for correctness. The practical applicability of the proposed framework is demonstrated using a case study of Air Traffic Control (ATC). ATC is a complex distributed safety-critical system, in which a minor error may cause catastrophic consequences leading to loss of human lives and heavy monetary penalties [39]. The mission of safe flight is distributed amongst several ground-based controllers, who monitor and control all aircrafts within their assigned air space.

In our case study, we propose a Metaverse application for the ATC system, in which an air traffic controller can perform its everyday aircraft controlling tasks from anywhere in the world after entering an integrated, immersive, and virtual world through a digital avatar. The proposed ATC Metaverse application is transformed into a formal model, to verify the ATC Metaverse application from different dimensions. The results of this integration help to analyse the correctness of the ATC Metaverse application.

The contributions of this article are threefold. First, the components of the Metaverse architecture are formally model using hierarchical CPN. It can serve as a template for mathematical verification of Metaverse applications and is available at GitHub [23]. The second contribution is a Metaverse application architecture of air traffic control (ATC), which will help air traffic controllers to effectively perform their duties from anywhere in the world. The third contribution is the procedure for increasing the level of confidence about the correctness of a formally verified model.

The outline of the rest of the paper is as follows: Section 2 gives the background knowledge, and Section 3 presents the architecture of the proposed extended Metaverse architecture. The formal modelling framework, which is called formal modelling and verification framework (FoMAV framework) for Metaverse applications, is presented in Section 4. Section 5 demonstrates a case study, showing how FoMAV is applied to formal modelling of an Air Traffic Control System in the Metaverse. In Section 6, an overall discussion is provided and, in Section 7, concluding remarks are enclosed.

## 2. Background and Research Gap

### *2.1.* Metaverse Components

Metaverse merges two words, 'meta' means beyond, and, 'verse' is the abbreviation of the universe. It impersonates the real world as a collection of 3D immersive virtual worlds connected via the Internet by using the concepts of virtual reality and augmented reality. With the increased demands of social media, the Metaverse is regarded as 3D Internet populated with people [47]. Therefore, the Metaverse is now considered as a future horizon, which is not something people may browse but rather a hypothetical universe to live in. Being an emerging technology the primary architecture of the Metaverse comprises of following seven layers as shown in Figure 1 [36].

*Layer 1 – Experience:* This layer represents the 3D environment of the physical world and the virtualization of real-time objects. It creates a hypothetical universe to live in. For example, the immersive experience, like shopping, theatre, games, social and e-sports.

*Layer 2 – Discovery:* It shows the discovery of new content and experiences. The discovery can be inbound, if the user wishes to search content himself, for instance, a shopping store, or game. It can be outbound, if the content is returned to the user for example advertisement, ratings, recommendations, and suggestions.

*Layer 3 – Creator Economy:* It allows creator to produce content without the knowledge of its implementation or programming. It also provides market buy and sale content. Other technologies include workflow and commerce.

*Layer 4 – Spatial Computing:* This layer allows a blend of the virtual and real world. The main technologies are virtual reality (VR), augmented reality (AR), the Cloud, sensors, and spatial mapping.

*Layer 5 – Decentralization:* It allows an open and distributed architecture, in which there is no central control. It allows the full command of creators on their data and contents and the backend services are reduced. Edge computing, AI agents, microservices, and blockchain are some important technologies of this layer.

*Layer 6 – Human Interface:* It allows transmission of information through touch and gestures without interaction with physical objects. This layer involves specialized hardware like smart glasses, wearables, haptic, gesture, voice, and neural sensor technologies.

*Layer 7 – Infrastructure:* This layer allows the development of the Metaverse. Being a backbone, this layer involves upgraded technologies like 6G, WiFi 6, Cloud, and graphics processing unit (GPU).

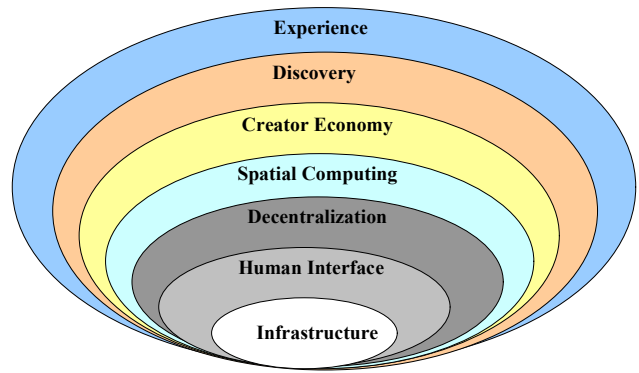

**Fig. 1.** Layers of Metaverse

## 2.2. Architecture of Metaverse

For this study, the capabilities of existing frameworks in the domain of the Metaverse are explored (Table 1). The research in [26] presented a conceptual architecture for the adoption of Metaverse in education institutes of gulf countries. Another study investigated the possible applications of the Metaverse in the field of education [10]. Svend et. al. proposed an architecture for Metaverse, focusing on its adoption in the marketing industry [52], and, in [41], a Metaverse architecture is provided for product retailers. Similarly, another research [31] presented a Metaverse architecture for the advertisement industry. John et. al. presented a survey on computing aspects of stand-alone 3D worlds, to facilitate their transitions into an integrated environment of Metaverse [28]. The research in [47] investigated networking and communicational challenges of an edge-enabled Metaverse, proposing three layers comprising infrastructure, Metaverse engine, and virtual world. Ning et. al. [21] proposed a technology-oriented architecture for the Metaverse. The research in [64] defined a multi-perspective architecture of the Metaverse and presented a discussion on opportunities, complexities, and challenges that are faced by organizations, users, and institutions.

These prior researches have limitations, in terms of their applicability in multiple domains. Either the Metaverse architectures pertain to a specific application, like education [10, 26], marketing [41, 52], advertisement [31], or focused on a specific point of interest, such as technology, security, or immersion (Table 1). No generalized architecture for multiple kinds of Metaverse application has been proposed so far.

**Table 1.** Comparison of proposed Metaverse architectures

| No. | Metaverse Domain | Contribution to Metaverse Architectures | Technique Applied | Reference |
|---|---|---|---|---|
| 1 | Metaverse architecture for educational purposes in the Gulf area | Presented a conceptual model that correlates personal characteristics with technology-based features | Deep learning, structural equation modelling, and technology acceptance models | [26] |
| 2 | Metaverse in education | Defined four types of Metaverse in education with respect to potentials and limitations | Classification of architectures | [10] |
| 3 | Evolution of the retail industry | Unveiled the need for policy studies for the Metaverse retail industry | Empirical analysis of "Second Life Economy" and proposed the immerse promotional strategies | [41] |
| 4 | Computing technologies that apply to 3D virtual spaces | Proposed the architectural direction for a scalable Metaverse. Outlined the need to move from a set of independent virtual worlds to an integrated network of 3D virtual worlds (Metaverse), constituting a compelling alternative for human sociocultural interaction | Survey and analysis of the existing technologies with respect to advancement towards Metaverse | [28] |
| 5 | Multidisciplinary, including marketing, education, and healthcare | Examined societal effects relating to social interaction in Metaverse, issues relating to trust, privacy, bias, disinformation, application of the law, psychological aspects linked to addiction, and impact on vulnerable people | Survey using informed narrative and multi-perspective approach from experts with varied disciplinary backgrounds | [64] |
| 6 | Marketing | Explained the concept of Metaverse as the 3D version of the Internet | Nike–Roblox case study, with which the customer benefits that are provided by the Nikeland project are explored | [52] |
| 7 | Technologies used in the Metaverse and its applications | Presented a technology-oriented Metaverse architecture | Literature Survey | [21] |
| 8 | Cloud-edge-enabled computation framework to realize the Metaverse | Presented an edge-enabled Metaverse architecture | Literature Survey | [47] |

## 2.3. Modelling of the Metaverse

Formal Methods are a set of rigorous mathematical techniques for the specification and verification of software and hardware systems. A system's formal model comprises concrete statements, which underlie firm foundations of well-defined disciplines, such as the set theory algebra and discrete calculus. Traditional informal approaches specify system requirements by using natural language or graphical notations [20]. The tremendous and diverse interpretation of natural language vocabulary makes informal specifications highly ambiguous for different stakeholders, which penetrate to later development phases as errors. Formal models, on the other hand, are composed of universally standardized notations of mathematics, which eliminate all potential ambiguities. Therefore, formal modelling enhances the understandability and precision of requirement specifications, giving deeper system insights [2].

Another powerful application of formal models is in the field of formal verification, which is a technique to verify the correctness of a developed formal model by executing sophisticated mathematical techniques such as model checking and deductive verifications [15]. The dynamic technique of simulation gets slow with an increase in model size, whereas formal verification is fast as it is based on static techniques that explore whether a formal model meets a specified property. Moreover, traditional informal system development techniques cannot execute requirement specifications. Therefore, potential errors are identified only during implementation and testing. Noteworthy, errors spotted during early phases of system development are easier and cheaper to be fixed [42]. Hence, formal modelling and formal verification lead to reliable and trustworthy system development.

There are seventy-nine notations, methods, and tools reported in the formal model virtual library [62]. They are broadly classified into two main categories: property-oriented and model-oriented [59]. In a property-oriented formal model, such as Algebraic Logic, Temporal Logic, a model is described as system operations and their relationships. In a model-oriented formal model, such as VDM, or Z, a system is specified by using states and transitions. Petri-nets are a widely used formal model for modelling dynamic and concurrent systems. The simple yet robust notations offered by the nets help to model complex data flows, such as concurrency conditional branching. They are also recommended for verification of critical areas, such as interdependencies, liveness, and reachability. Therefore, they are successfully applied for the formal modelling of complex systems, like manufacturing systems, communication protocols, and AI systems [58].

The dominating features offered by formal modelling have excelled in its widespread adoption in both academia and industry. There is an increasing trend in the emergence of improved formal languages, effective verification approaches, and academic recognition in form of awards [3]. Many very large businesses, like Microsoft, NASA, and Facebook, have converged on using formal methods for the specification and verification of both software and hardware systems [22]. Studies have demonstrated the benefits of applying formal modelling to complex safety-critical systems [33, 34, 35].

However, there is a lack of discussion about the formalization of the Metaverse. There are only some stand-alone formal modelling and verification, irrespective of Metaverse, for certain components like AI [49, 50, 63], ubiquitous systems [1, 27, 45], cloud and edge computing [3, 4, 5, 7, 42, 61], and blockchain [57, 60, 65].

Table 2 presents the availability of formal specifications in the literature for different Metaverse layers along with their supporting technologies as defined in [47]. The fourth column of Table 2 is filled with the notation "Y", "N", and "Y*". If the formal specification of the relevant "Metaverse Layer" (column 2 of Table 2) and the "Supporting Technology" (column-3 of Table-2) is available in the literature, then it is represented with 'Y'. 'N' means the technology has not been formally specified so far. Whereas 'Y*' shows that no explicit formal specification exists, but its formalization is an implicit part of some other related domain.

**Table 2.** Literature review of formal specifications of supporting technologies of different Metaverse layers

| No. | Metaverse Layers | Supporting Technology | Formal Specifications | Reference |
|---|---|---|---|---|
| 1 | Experience | Smart Hyper-personalized Games | N | - |
| | | Personal Experiences | N | - |
| | | Education | Y* | [56] |
| | | Welfare Assistant | Y* | [38] |
| 2 | Discovery | Intelligent Network | Y | [6] |
| 3 | Creator Economy | Design Tools | N | - |
| | | Asset Market | Y | [46] |
| | | Workflow | Y | [11] |
| | | Commerce | N | - |
| 4 | Spatial Computing | 3D Engines | Y | [44] |
| | | VR/AR/XR | Y | [30] |
| | | Multitasking UI | Y* | [37] |
| | | Geospatial Mapping | Y | [66] |
| 5 | Decentralization | Edge Computing | Y | [13] |
| | | AI agent | Y | [48] |
| | | Microservices | Y | [16] |
| | | Blockchain | Y | [51] |
| 6 | Human Interface | Touch | N | - |
| | | Gesture | N | - |
| | | Voice | Y* | [53] |
| | | Neural | Y* | [50] |
| 7 | Infrastructure | 6G | N | - |
| | | WiFi 6 | N | - |
| | | Cloud | Y | [9] |
| | | Micro-electromechanical systems | Y* | [54] |
| | | Graphics processing units | N | - |

**2.4. Formal Verification**

Finally, we survey the literature about formal verification of the Air Traffic Control (ATC) system, which is a very focused research area. The application of formal modelling in the design phase of safety-critical systems helps to reveal potential subtle errors leading to increased correctness, consistency, and accuracy of the system. The research in [4] provided formal modelling (FM) and verification of an ATC assistance system for error identification. The work in [3] applied FM in ATC scenarios for task load and resource conflicts. In [35], a formal model of the aircraft take-off procedure is presented for safety analysis. However, the implementation of ATC in Metaverse technology is not available in the literature. Only one Metaverse for aircraft maintenance was found in the literature, but it is just a media player [8] for the training of aircraft maintenance. Some ATC games, such as HTML5 have claimed of implementing the Metaverse technology.

There is a growing discussion on Metaverse in future air transportation. According to Reuters [25], "Boeing" is endeavouring to build its next aircraft by using Metaverse technology. Frequentis [24] is a safety-critical communication and information solution envisioned as a completely new approach to controllers of the tower operations centre, driven by the combination of modern and digital immersive technologies. However, it is not based on virtual reality (VR).

**2.5. Research Gap**

Based on the literature review presented in subsections 2.2, 2.3, and 2.4, the following two research gaps can be identified. Firstly, the prior studies on Metaverse architectures have certain limitations in terms of their applicability in multiple domains (Subsection 2.2). Either, there are Metaverse architectures pertaining to a specific application, such as education [10, 26], marketing [41, 52], and advertisement [31], or the Metaverse architecture focused only on a specific point of interest such as technology, security, or immersion (Table 1). Moreover, there is a lack of formal modelling and verification of each constituent component of the Metaverse (subsection 2.4), and there is no integrated approach for the application of formal modelling in the Metaverse. This situation is understandable, as Metaverse is a complex integration of interoperable, immersive, and un-divided (shardless) virtual 3D environments, and its modelling is a challenging task. Consequently, we can formulate the first research question as whether a Metaverse architecture can be developed that can be applied to a large set of applications.

Secondly, Table 2 shows the availability of formal modelling for specific Metaverse applications. Based on the literature review in Subsection 2.3, it can be concluded that there is no generalized framework for the formal modelling of diverse Metaverse applications. Keeping in view that the Metaverse is an emerging adaptation of the Internet and many organizations already started to assess its potential for integrating it within their existing operations and business models, it becomes clear that there is a need to evaluate various dimensions of emerging Metaverse technologies before integrating it into real-world applications. It becomes clear that the lack of a generalized framework is a significant shortcoming. Consequently, we can formulate the second research question as whether it is possible to develop a flexible formal modelling framework for diverse Metaverse applications.

**3. Proposed Architecture of the Metaverse**

Based on the literature review in Subsection 2.2, an architecture for integrating Metaverse technologies is proposed. Our work extends the architecture presented in [47], which focuses on technologies related to edge computing. It rigorously identifies and integrates components, so that the resulting architecture fulfils the requirements of every Metaverse application. Figure 2 shows our developed Metaverse architecture, which is composed of four layers and two pillars.

The infrastructure layer is the core layer of the Metaverse and provides basic functionalities of communication, computation, and storage. Above it, the Metaverse engine layer exists, which gives fundamental services such as AR/VR, Haptic, Digital Twin, AI, and Blockchain. The virtual world layer resides as a third layer, which represents immersive features, like Avatars, Virtual Environments, Virtual Goods / Services, and Tangible Goods / Services. The fourth, topmost layer is the Physical Layer, which depicts tangible components from the real world such as the user, IoT, sensors, virtual service providers, and physical service providers.

Our work also enhances the existing Metaverse architecture through the identification of two new components that are not integrated into earlier studies, namely formal modelling and formal verification as well as security and privacy. As these two components are linked with all four basic layers of the proposed Metaverse architecture, they are called pillars. The formal modelling and verification pillar provide services for mathematically specifying every Metaverse layer of the proposed framework. These specifications can be used for analysing the correctness, by which undesired system behaviour and potential subtle errors may be identified. Similarly, the security and privacy pillar provide the features of restricting unauthorized access to Metaverse data and user information, preventing cybercrimes and other cyberthreats. The following sub-sections cover details of each layer and pillar of the proposed Metaverse architecture with their associated components.

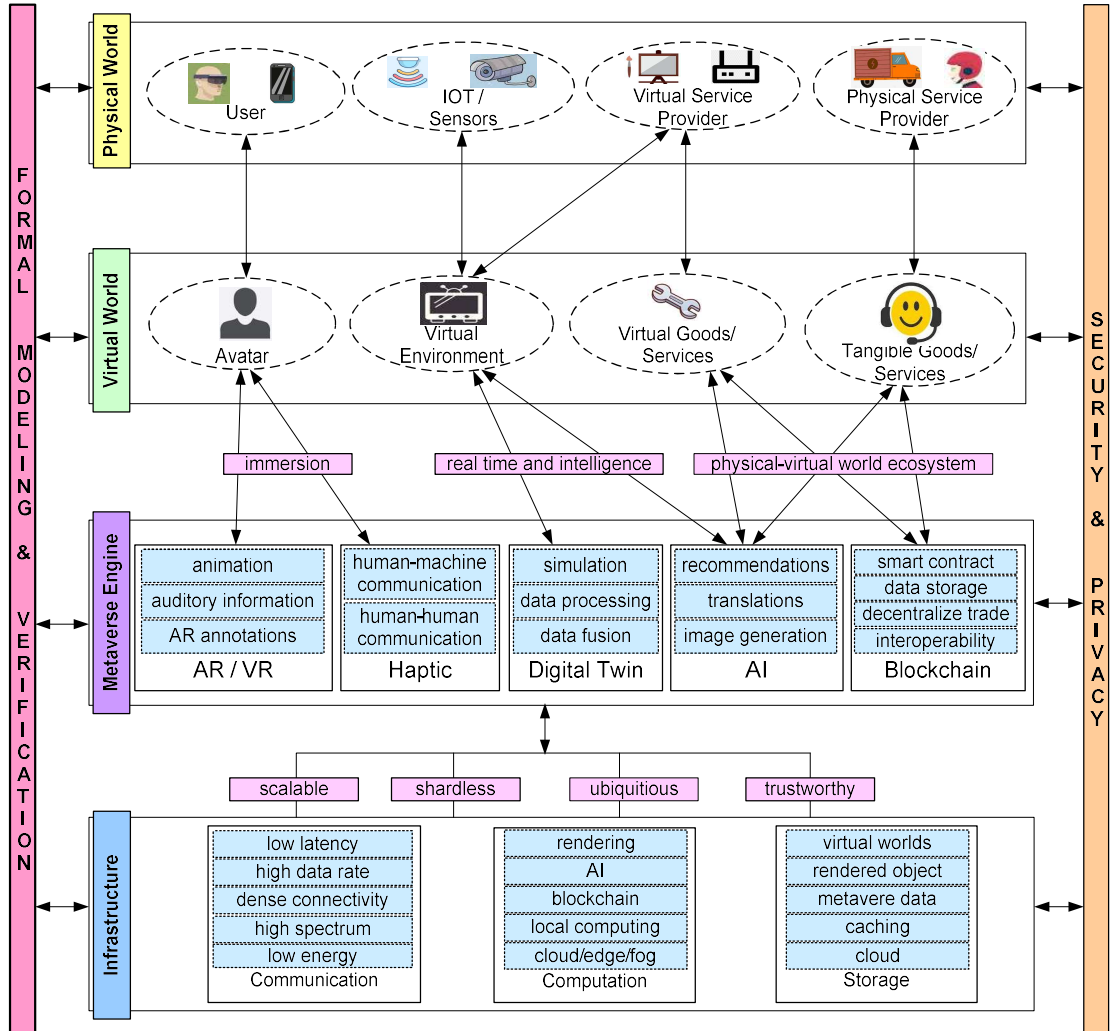

**Fig. 2.** Proposed architecture of Metaverse

### 3.1. Infrastructure Layer

The infrastructure layer is the core layer of the Metaverse and provides basic functionalities of communication, computation, and storage. Computation is an extensive set of services offered by Metaverse from data gathering to rendering data and videos for providing an immersive experience to users that require expensive computations to be performed by processors and GPUs. The four main computation-intensive processes of Metaverse are rendering, AI, blockchain, and local computing. In Metaverse, the display for users applies rendering, which is the process of transforming raw data into 3D images and videos. Artificial Intelligence is the backbone of the Metaverse, which is used for the creation of immersive experiences, AR/VR applications, and gesture recognitions. For the low resource-consuming processes, local computations are performed, whereas, for high-consuming processes, technologies like cloud, edge, and fog computing are applied for distributing the load among the devices. The allocation of tasks to the different resources is determined based on optimization algorithms.

Metaverse is a complex network and requires a large amount of storage. There are three types of storage required by a Metaverse. The need for storage increases rapidly depending on what task the user tries to accomplish through the Metaverse application. First, there is need for storage required by the overall Metaverse world state itself, which involves the virtual worlds generated on the user's demand. Second, it is important to store the digital assets, such as, rendered objects produced during the execution of Metaverse applications. Third, there is a need to store the data of users generated by their interaction with the Metaverse.

There is an ultra-high-speed data communication need of the Metaverse for virtualization and haptic commands. New emerging technologies, like B5G (Beyond 5G also called 6G), must be integrable. The communication medium of Metaverse requires low latency, high data rate, dense connectivity, high spectrum, and low energy demands.

### 3.2. Metaverse Engine Layer

Above the infrastructure layer, lays the Metaverse engine layer, which gives fundamental services such as AR/VR, Haptic, Digital Twin, AI, and Blockchain. VR/AR are the important and primary concepts of Metaverse technology. VR creates an immersive experience for a user by transforming real-world images and objects into a computer-generated environment. AR focuses on the real-time integration of text, graphics, audio, and other virtual enhancements with real-world objects. In our proposed framework, it is defined with the characteristics like animation, auditory information, and AR annotations. The Haptic technology allows for sending and interpreting information through the sense of touch. The proposed framework of Metaverse paid attention to the human to machine and human-to-human communication by using haptic.

A digital twin is an implementation of a real-world entity such as a person, object, or organization into its digital representation. In the proposed Metaverse framework, a digital twin requires simulation, data processing, and data fusion. AI is the primary feature defined in the Metaverse engine layer (Figure 2). It empowers a computing device to solve problems and make decisions just like a human brain. The recommendations, translations, and image-generation capabilities of AI are leveraged in our proposed Metaverse architecture. Blockchain is the process of storing transactions and tracing assets in the form of a shared and immutable ledger. This interesting new feature of the Metaverse is defined by features like smart contracts, data storage, decentralized trading, and data interoperability.

### 3.3. Virtual World Layer

The third layer of the proposed framework is composed of the following components: avatar, virtual environment, virtual goods/services and physical goods/services. An avatar is a customizable physical representation of the user in a computer-generated 3D world that can be used primarily in chat and entertainment websites. In Metaverse, a virtual environment is a user-created immersive editable digital 3D world, needed to socialize with other people, participate in an event and gain access or experience to any real-world venue. Virtual and physical goods/services are defined as intangible and tangible assets, respectively. They can be traded in a virtual online economy, in which their costs are determined by offers of buyers.

### 3.4. Physical World Layer

The user can enter the Metaverse through headsets, smart goggles, cell phones, and hand-operated controllers. Internet-of-Things (IoT) devices connect physical objects, in order to interact and share data with other objects or systems by using sensors technologies via the Internet. Virtual and physical service providers are defined as buyers or sellers of intangible and tangible assets, respectively. Those can be traded in a virtual online economy.

### 3.5. Formal Modeling and Verification Pillar

Formal modelling being an important pillar of our Metaverse architecture is needed in all other Metaverse components, which means that all layers and their corresponding components need formal modelling and verification. Metaverse is a complex domain and the application of formal modelling in it provides a precise and unambiguous approach to requirements specification. In this way, a deeper system understanding is achieved, also critical issues, as well as misconceptions, are identified during the system's mathematical specifications. Metaverse integrates four interacting layers and any incorrect or abnormal system behaviour may lead to difficult and fatal runtime errors. Through formal verification, in the Metaverse, such critical issues are identified during the system's modelling phase, when bug identification and removal is easier and cheaper. Moreover, formal modelling can be applied in the domain of Metaverse during any phase of the system development lifecycle, in order to obtain the aforementioned advantages. Therefore, the application of formal modelling in the Metaverse leads to the development of high-quality and reliable applications. The details of the formal modelling and verification pillar can be detailed with a formal modelling and verification framework (Section 4).

### 3.6. Security and Privacy Pillar

In our proposed Metaverse architecture, the security and privacy pillar is interconnected with every layer. Metaverse provides a digitally immersive experience to its users and ranges its applications in education, healthcare, real estate, decentralized trading, and marketing. Such diversity in the Metaverse demands ultra-high security and privacy. The privacy of data related to user profiles, individual participation, access, and digital goods/assets such as cryptocurrency and virtual services/applications are at high risk. Similarly, some important security issues pertaining to Metaverse technology are identity theft, stealing virtual assets, malware attacks, harassment, and cyberbullying. Therefore, security and privacy features need to be applied on each layer of the proposed Metaverse architecture and their constituting components.

## 4. Formal Model and Verification Framework (FoMAV)

After formulation of a generic Metaverse framework, the interconnections of the Formal Model and Verification Pillar with the other layers need to be defined. Although the formalization of specific, stand-alone components exists, the challenge is the potential applicability of a single formal modelling method to every constituent module of the proposed generic Metaverse framework. However, managing the complex interactions within the Metaverse may require numerous state and transition definitions, leading to models that are difficult to manage.

Eventually, the hierarchical CPN were identified as a potential solution. Petri nets are widely applied for modelling dynamic and concurrent systems [32]. They are suitable for formal modelling of the proposed framework, since the Metaverse is an integration of several distributed concurrent units. The application of Petri nets has successfully demonstrated the modelling of some complex systems, such as manufacturing systems, communication protocols, and AI systems [58]. The selected variant of Petri nets enhances the traditional framework by incorporating user-defined data types as token colours, increasing expressiveness. CPNs are versatile tools for modelling and analysing complex, concurrent, and dynamic systems, offering graphical representation and formal semantics for precise analysis and verification, making them suitable for various domains, including safety-critical systems in aviation, such as aircraft control and air traffic procedures. Additionally, hierarchical CPN enable the structuring of the proposed model into sub-models, reducing system complexity.

### 4.1. Data Type

We have defined a new data type named PACKET for the formal modelling of Metaverse, in order to represent a data unit to be transmitted and received by the Metaverse layers. In CPN, "colour" refers to a data type that allows tokens to carry additional information, enabling more detailed and expressive modelling of complex systems.

A PACKET is mathematically declared as a tuple of STRING (a sequence of characters) to represent the receiver, a STRING (a sequence of characters) to depict data, and an INT (a whole number) to show flag value. Generally, a flag represents a value indicating a condition or status within a program. In our work, a "flag value" is a specific numerical identifier used to denote a command requested by the sender, aiding in accurate communication and execution within the system.

For example, when a packet is sent from a virtual environment to a Metaverse Engine (ME), the token values are set to ("ME", "DATA", 3), where the receiver field is set as "ME", DATA is represented as an abstract type, and the flag value of 3 represents the number associated with a command requested by the sender. In a hierarchical CPN, a model can be structured into a number of coordinating Petri nets called subnets. A subnet can receive input from other subnets through a special type of place, which is called an input port. A set of input places is defined as $Port_{in}$. For example, $Port_{in} : P \rightarrow$ subnet is a function from the set of input places to the subnet which sends input. Similarly, a subnet can send output to other subnets through a special type of place called output port. A set of output places is defined as $Port_{out}$. For example, $Port_{out} : P \rightarrow$ subnet is a function from the set of output places to the subnet, which requires the output. Since the formal model acts as a template for all Metaverse applications, the components of a layer can be enhanced or removed depending on the type of application to be modelled. Each layer of the proposed Metaverse is formalized into a Petri net subnet, the details are defined as follows.

### 4.2. Modelling of Interconnection of Metaverse Layers

The architectural diagram of the main hierarchical Petri net is shown in Figure 3. It is showing the interconnection among all coordinating subnets. We have developed Petri net subnets for each of the four basic Metaverse layers named Physical World, Virtual World, Metaverse Engine and Infrastructure as defined in the proposed Metaverse framework (Section 3). The interrelationships between the subnets are defined in the same way as shown in the proposed framework. Each subnet is defined as a tuple $C = (\Sigma, P, T, A, N, C, G, E, I)$, which comprises elements as defined in Table 3. The details of the formalization of each Metaverse layer into Petri nets subnets are provided in the following subsections.

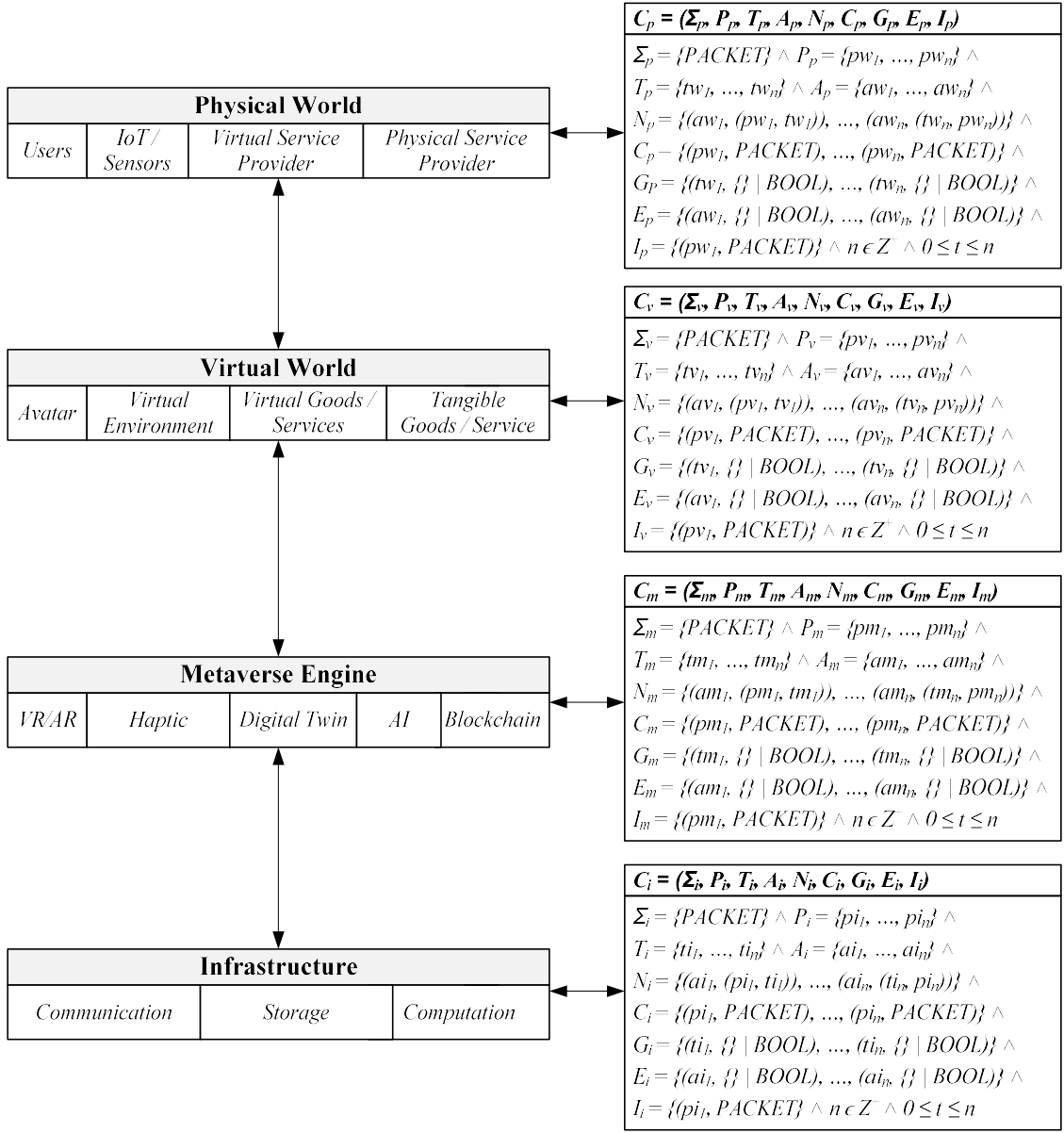

**Fig. 3.** Hierarchical Representation of Proposed Metaverse Architecture

**Table 3.** Description of elements defined for modelling the proposed Metaverse architecture

| Element | Description | Element | Description |
|---|---|---|---|
| Σ | Set of user-defined data types | C | Function to map places with their colours |
| P | Set of places | G | Guard Function to attach transitions with Boolean condition |
| T | Set of transitions | E | Arc expression function to map each arc with a Boolean condition |
| A | Set of arcs | I | A binary relation to define token for initial place |
| N | Function to bind arcs with their endpoints | - | - |

### 4.3. Petri-Net Subnet Representing the Physical World Layer

The physical world layer is represented by a subnet named Physical World, which is mathematically defined as a tuple, $C_p$.

$$C_p = (\Sigma_p, P_p, T_p, A_p, N_p, C_p, G_p, E_p, I_p)$$
$$\Sigma_p = \{PACKET\} \land P_p = \{pw_1, ..., pw_n\} \land$$
$$T_p = \{tw_1, ..., tw_n\} \land A_p = \{aw_1, ..., aw_n\} \land$$
$$N_p = \{(aw_1, (pw_1, tw_1)), ..., (aw_n, (tw_n, pw_n))\} \land$$
$$C_p = \{(pw_1, PACKET), ..., (pw_n, PACKET)\} \land$$
$$G_P = \{(tw_1, \{\} \mid BOOL), ..., (tw_n, \{\} \mid BOOL)\} \land$$
$$E_p = \{(aw_1, \{\} \mid BOOL), ..., (aw_n, \{\} \mid BOOL)\} \land$$
$$I_p = \{(pw_1, PACKET)\} \land n \in Z^+ \land 0 \leq t \leq n$$

where the set of colours is $\Sigma_p$, the set of places $P_p$, the set of transitions $T_p$, the set of all connecting arcs $A_p$, the node function $N_p$ to bind each arc with its endpoints, the colour function $C_p$ to map each place with a colour, the guard function $G_P$ to attach each transition with a guard condition, the arc expression function $E_p$ to map each arc with an arc expression and the initialization function $I_p$ to define token for the initial place, where n is a positive integer which can be between 0 and n, and t is the number of tokens ranges from 0 to n. This subnet comprises four other subnets, which are mathematically specified as follows.

**User**

It is a subset of Physical World, $C_p$, and there are places p1 and p2, in which p1 is an input port for avatar of subnet Virtual World and p2 is an output port for avatar of subnet Virtual World abbreviated as VW.

$$User \subset C_p$$
$$p1, p2 \in P_p \land$$
$$p1 \in Port_{in} \land p1.Subnet = Avatar.VW \land$$
$$p2 \in Port_{out} \land p2.Subnet = Avatar.VW$$

**Internet of Things (IOT)/Sensor**

It is a subset of the Physical World, $C_p$, and there are places p1 and p2, in which p1 is an input port for the Virtual Environment of the subnet Virtual World and p2 is an output port for the Virtual Environment of the Virtual World (VW).

| *IoT / Sensor* $\subset C_p$ |
|---|
| $p1, p2 \in P_p \wedge$ |
| $p1 \in Port_{in} \wedge p1.Subnet = Virtual\ Environment.VW \wedge$ |
| $p2 \in Port_{out} \wedge p2.Subnet = Virtual\ Environment.VW$ |

**Virtual Service Provider**

It is a subset of Physical World, $C_p$, and there are places p1, p2, which are input ports for Virtual Environment and Virtual Goods/Services of subnet Virtual World, respectively. Also, there are places p3 and p4, which are output ports for Virtual Environment and Virtual Goods/Services of subnet Virtual World, respectively.

| *Virtual Service Provider* $\subset C_p$ |
|---|
| $p1, p2, p3, p4 \in P_p \wedge$ |
| $p1, p2 \in Port_{in} \wedge p1.Subnet = Virtual\ Environment.VW \wedge$ |
| $\qquad p2.Subnet = Virtual\ Good/Services.VW \wedge$ |
| $p3, p4 \in Port_{out} \wedge p3.Subnet = Virtual\ Environment.VW \wedge$ |
| $\qquad p4.Subnet = Virtual\ Good/Services.VW$ |

**Physical service provider**

It is a subset of Physical World, $C_p$, and there are places p1, p2, in which p1 is an input port for Tangible Goods/Services of Virtual World and p2 is an output port for Tangible Goods/Services of Virtual World (VW).

| *Physical Service Provider* $\subset C_p$ |
|---|
| $p1, p2 \in P_p \wedge$ |
| $p1 \in Port_{in} \wedge p1.Subnet = Tangible\ Goods\ /\ Services.VW \wedge$ |
| $p2 \in Port_{out} \wedge p2.Subnet = Tangible\ Goods\ /\ Services.VW$ |

**4.4. Petri-Net Subnet Representing the Virtual World Layer**

The virtual world layer is represented by a subnet named Virtual World, which is mathematically defined as a tuple, $C_v$, where the set of colors is $\Sigma_v$, the set of places $P_v$, the set of transitions $T_v$, the set of all connecting arcs $A_v$, the node function to bind each arc with its endpoints $N_v$, the color function to map each place with a color $C_v$, the guard function to attach each transition with a guard condition $G_v$, the arc expression function to map each arc with an arc expression $E_v$, and the initialization function to define token for the initial place, $I_v$, where n is a positive integer and t, which ranges from 0 to n, is the number of tokens. This subnet comprises four other subnets, which are mathematically specified as follows.

| $C_v = (\Sigma_v, P_v, T_v, A_v, N_v, C_v, G_v, E_v, I_v)$ |
|---|
| $\Sigma_v = \{PACKET\} \wedge P_v = \{pv_1, ..., pv_n\} \wedge$ |
| $T_v = \{tv_1, ..., tv_n\} \wedge A_v = \{av_1, ..., av_n\} \wedge$ |
| $N_v = \{(av_1, (pv_1, tv_1)), ..., (av_n, (tv_n, pv_n))\} \wedge$ |
| $C_v = \{(pv_1, PACKET), ..., (pv_n, PACKET)\} \wedge$ |
| $G_v = \{(tv_1, \{\} \mid BOOL), ..., (tv_n, \{\} \mid BOOL)\} \wedge$ |
| $E_v = \{(av_1, \{\} \mid BOOL), ..., (av_n, \{\} \mid BOOL)\} \wedge$ |
| $I_v = \{(pv_1, PACKET)\} \wedge n \in Z^+ \wedge 0 \leq t \leq n$ |

**Avatar**

It is a subset of Virtual World, $C_p$ and there is a place p1, which is an input port for the User of Physical World. Also, there are places p3 and p4, which are input ports for Virtual Reality/Augmented Reality and Haptic of Metaverse Engine respectively. There is a place p2, which is the output port for the User of Virtual World.

| *Avatar* $\subset C_v$ |
|---|
| $p1, p2, p3, p4 \in P_v \wedge$ |
| $p1, p3, p4 \in Port_{in} \wedge p1.Subnet = User.PW \wedge$ |
| $\quad\quad p3.Subnet = Virtual\ Reality/Augmented\ Reality.ME \wedge$ |
| $\quad\quad p4.Subnet = Haptic.ME \wedge$ |
| $p2 \in Port_{out} \wedge p2.Subnet = User.PW$ |

**Virtual Environment**

It is a subset of Virtual World, $C_v$, and there are places p1 and p2, which are input ports for IoT/Sensor and Virtual Service Provider of Physical World, respectively. Also, there are places p3 and p4, which are output ports for IoT/Sensor and Virtual Service of Physical World, respectively. There are places p5 and p6, which are input ports for Artificial Intelligence and Digital Twin of Metaverse Engine respectively. PW stands for the physical world, and ME stands for Metaverse Engine.

| *Virtual Environment* $\subset C_v$ |
|---|
| $p1, p2, p3, p4, p5, p6 \in P_v \wedge$ |
| $p1, p2, p5, p6 \in Port_{in} \wedge p1.Subnet = IOT/Sensor.PW \wedge$ |
| $\quad\quad p2.Subnet = Virtual\ Service\ Provider.PW \wedge$ |
| $\quad\quad p5.Subnet = Artificial\ Intelligence.ME \wedge$ |
| $\quad\quad p6.Subnet = Digital\ Twin.ME \wedge$ |
| $p3, p4 \in Port_{out} \wedge p3.Subnet = IOT/Sensor.PW \wedge$ |
| $\quad\quad p4.Subnet = Virtual\ Service\ Provider.PW$ |

**Virtual Goods/Services**

It is a subset of Virtual World, $C_v$, and there is a place p1, which is an input port for the Virtual Service Provider of subnet Physical World. Also, there are places p3 and p4, which are input ports for Artificial Intelligence and Blockchain of subnet Metaverse Engine, respectively. There is a place p2, which is an output port for the Virtual Service Provider of subnet Physical World. PW stands for Physical World, and ME stands for Metaverse Engine.

| *Virtual Goods/Services* $\subset C_v$ |
|---|
| $p1, p2, p3, p4 \in P_v \wedge$ |
| $p1, p3, p4 \in Port_{in} \wedge p1.Subnet = Virtual\ Service\ Provider.PW \wedge$ |
| $\quad\quad p3.Subnet = Artificial\ Intelligence.ME \wedge$ |
| $\quad\quad p4.Subnet = Blockchain.ME \wedge$ |
| $p2 \in Port_{out} \wedge p2.Subnet = Virtual\ Service\ Provider.PW$ |

**Tangible Goods / Services**

It is a subset of Virtual World, $C_v$, and there is a place p1, which is an input port for the Physical Service Provider of subnet Physical World. Also, there are places p3 and p4, which are input ports for Artificial Intelligence and Blockchain of subnet Metaverse Engine respectively. There is a place p2, which is an output port for the Physical Service Provider of subnet Physical World, where PW stands for the Physical World and ME stands for Metaverse Engine.

| *Tangible Goods/Services* ⊂ $C_v$ |
|---|
| $p1, p2, p3, p4 \in P_v \land$ |
| $p1, p3, p4 \in Port_{in} \land p1.Subnet = Physical\ Service\ Provider.PW \land$ |
| $\quad p3.Subnet = Artificial\ Intelligence.ME \land$ |
| $\quad p4.Subnet = Blockchain.ME \land$ |
| $p2 \in Port_{out} \land p2.Subnet = Physical\ Service\ Provider.PW$ |

**4.5. Petri-Net Subnet Representing the Metaverse Engine Layer**

The Metaverse engine layer is represented by a subnet named Metaverse Engine, which is mathematically defined as a tuple, $C_m$.

| $C_m = (\Sigma_m, P_m, T_m, A_m, N_m, C_m, G_m, E_m, I_m)$ |
|---|
| $\Sigma_m = \{PACKET\} \land P_m = \{pm_1, ..., pm_n\} \land$ |
| $T_m = \{tm_1, ..., tm_n\} \land A_m = \{am_1, ..., am_n\} \land$ |
| $N_m = \{(am_1, (pm_1, tm_1)), ..., (am_n, (tm_n, pm_n))\} \land$ |
| $C_m = \{(pm_1, PACKET), ..., (pm_n, PACKET)\} \land$ |
| $G_m = \{(tm_1, \{\} \mid BOOL), ..., (tm_n, \{\} \mid BOOL)\} \land$ |
| $E_m = \{(am_1, \{\} \mid BOOL), ..., (am_n, \{\} \mid BOOL)\} \land$ |
| $I_m = \{(pm_1, PACKET)\} \land n \in Z^+ \land 0 \leq t \leq n$ |

where the set of colours is $\Sigma_m$, the set of places $P_m$, the set of transitions $T_m$, the set of all connecting arcs $A_m$, the node function $N_m$ to bind each arc with its endpoints, the colour function $C_m$ to map each place with a colour, the guard function $G_m$ to attach each transition with a guard condition, the arc expression function $E_m$ to map each arc with an arc expression, and the initialization function $I_m$ to define token for the initial place, where n is a positive integer and t is the number of tokens that can range from 0 to n. This subnet comprises of four other subnets, which are mathematically specified as follows.

**Virtual Reality/Augmented Reality**

It is a subset of Metaverse Engine, $C_m$, and there is a place p, which is an output port for Avatar of Virtual World (VW).

| *Virtual Reality/Augmented Reality* ⊂ $C_m$ |
|---|
| $p \in P_m \land$ |
| $p \in Port_{out} \land p.Subnet = Avatar.VW$ |

**Haptic**

It is a subset of Metaverse Engine, $C_m$, and there is a place p, which is an output port for avatar of Virtual World (VW).

| *Haptic* ⊂ $C_m$ |
|---|
| $p \in P_m \land$ |
| $p \in Port_{out} \land p.Subnet = Avatar.VW$ |

**Digital Twin**

It is a subset of Metaverse Engine, $C_m$, and there a place p, which is an output port for Virtual Environment of Virtual World (VW).

| *Digital Twin* ⊂ $C_m$ |
|---|
| $p \in P_m \land$ |
| $p \in Port_{out} \land p.Subnet = Virtual\ Environment.VW$ |

**Artificial Intelligence**

It is a subset of Metaverse Engine, $C_m$, and there are places p1, p2, and p3, which are output ports for Virtual Environment, Virtual Goods/Services, and Tangible Goods/Services of Virtual World (VW).

> *Artificial Intelligence $\subset C_m$*
> *p1, p2, p3 $\epsilon$ $P_m$ $\wedge$*
> *p1, p2, p3 $\epsilon$ $Port_{out}$ $\wedge$ p1.Subnet = Virtual Environment.VW $\wedge$*
> *p2.Subnet = Virtual Goods/Services.VW $\wedge$*
> *p3.Subnet = Tangible Goods/Services.VW*

**Blockchain**

It is a subset of Metaverse Engine, $C_m$, and there are places p1 and p2, which are output ports for Virtual Goods/Services and Tangible Goods/Services of Virtual World (VW).

> *Blockchain $\subset C_m$*
> *p1, p2 $\epsilon$ $P_m$ $\wedge$*
> *p1, p2 $\epsilon$ $Port_{out}$ $\wedge$ p1.Subnet = Virtual Goods/Services.VW $\wedge$*
> *p2.Subnet = Tangible Goods/Services.VW*

### 4.6. Petri-Net Subnet Representing the Infrastructure Layer

The Infrastructure layer is represented by a subnet named Infrastructure, which is mathematically defined as a tuple, $C_i$.

> *$C_i = (\Sigma_i, P_i, T_i, A_i, N_i, C_i, G_i, E_i, I_i)$*
> *$\Sigma_i = \{PACKET\} \wedge P_i = \{pi_1, ..., pi_n\} \wedge$*
> *$T_i = \{ti_1, ..., ti_n\} \wedge A_i = \{ai_1, ..., ai_n\} \wedge$*
> *$N_i = \{(ai_1, (pi_1, ti_1)), ..., (ai_n, (ti_n, pi_n))\} \wedge$*
> *$C_i = \{(pi_1, PACKET), ..., (pi_n, PACKET)\} \wedge$*
> *$G_i = \{(ti_1, \{\} \mid BOOL), ..., (ti_n, \{\} \mid BOOL)\} \wedge$*
> *$E_i = \{(ai_1, \{\} \mid BOOL), ..., (ai_n, \{\} \mid BOOL)\} \wedge$*
> *$I_i = \{(pi_1, PACKET\} \wedge n \epsilon Z^+ \wedge 0 \leq t \leq n$*

where the set of colours is $\Sigma_i$, the set of places $P_i$, the set of transitions $T_i$, the set of all connecting arcs $A_i$, the node function $N_i$ to bind each arc with its endpoints, the colour function $C_i$ to map each place with a colour, the guard function $G_i$ to attach each transition with a guard condition, the arc expression function $E_i$ to map each arc with an arc expression, and the initialization function $I_i$ to define token for initial place, where n is a positive integer and t is the number of tokens those can be from 0 to n. This subnet comprises of three subnets, which are defined as abstract data types named Communication, Computation, and Storage. They are defined as abstract data types in order to hide unnecessary details which reduces model complexity and increases efficiency.

## 5. Formal Modelling of Air Traffic Control System in Proposed Metaverse Framework (FoMAV)– A Case Study

In this section, the practical applicability of the proposed Metaverse framework is demonstrated using the case study of the Air Traffic Control (ATC) system. It is a complex distributed safety-critical system, in which even a minor error may cause catastrophic consequences leading to loss of human lives and heavy monetary penalties [39]. The services offered to an aircraft by ATC controllers are still not fully computerized [5]. This restricts ground-based controllers to perform their flight safety duties and have to sit physically in airport towers to monitor and control all aircrafts within their assigned air space. We propose a Metaverse architecture for the ATC system (Figure 4), by which an ATC controller can perform its everyday aircraft controlling tasks from anywhere in the world, once she/he enters an integrated, immersive, and virtual world through a digital avatar. This Metaverse is derived from the formally integrated Metaverse framework described above. Some components of the layers of the proposed Metaverse framework, like blockchain, and physical goods/services are not included in this application, because their services are not required.

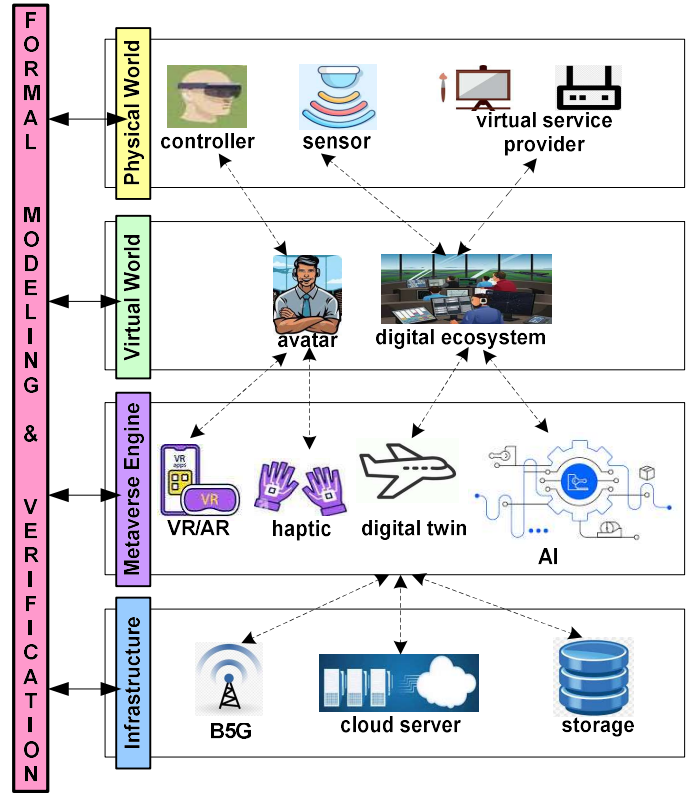

**Fig. 4.** Conceptual diagram of air traffic control (ATC) Metaverse, which has been derived from the Metaverse framework.

Starting from services of the top most layer named the "Physical Layer" of the Metaverse, which primarily provides integrated access to a digital world for **ATC controllers** through its headsets. **Sensors** data from the real world is obtained and the "**Virtual Service Provider**" gives necessary support for the virtualization of the real world, through which the ATC controller can perform its everyday tasks. Then, the components of the Virtual World layer come into play by creating **avatars** of a controller and generating an immersive virtual **digital ecosystem** of a physical airport tower, in which a controller actually works and perform its aircraft monitoring and control duties. The third Metaverse layer called the Metaverse engine actuates its components and, through the application of **VR/AR,** the **digital twins** of aircrafts to be supervised by the controller's avatar are produced. The users, who are the ATC controllers, can issue movement commands to traverse the digital ecosystem and also issue three main instructions through **haptic**. Local computation is performed by the Metaverse engine to issue guidance to aircrafts, while seeking sensor inputs from the physical world. The controller can get recommendations about conflict resolution by utilizing **AI** components. There are many conflicting situations, in which a pilot seeks ATC guidance, and some of these are summarized in Table 4. The details are available in [35]. The infrastructure layer provides communication through high-speed Internet called **B5G**, **cloud services** are utilized for computation, in order to carry out complex intensive computations, and **storage** provides placeholders for Metaverse creations and caches also safety-related tutorials that can be retrieved by the ATC controller.

**Table 4.** Aircraft safety conflicts and recommendations.

| Scenario | Conflict | Recommendations |
|---|---|---|
| 1 | Aircrafts trying to land at the same time | Select one aircraft and delay other |
| 2 | One aircraft is taking off and another one is landing | Assign alternate runway |
| 3 | An aircraft is de-trailing and another one is landing | Assign alternate runway |

There are many conflicting situations in which a pilot seeks ATC guidance and some of them are summarized in Table 4, the details are available in [35]. The second column of Table 4 shows conflicting situations detected by sensors from physical world and third column shows the recommendations. In row 1 the conflict describes a scenario in which two aircrafts are trying to land at same time which is recommended to be resolved by the technique to select one aircraft for landing and delay other one. Second scenario described in row 2 shows a scenario in which one aircraft is scheduled to take off and another landing at the same time which is recommended to be resolved by the technique to assign current runway to one and an alternate runway to the other aircraft. In row 3 the conflict describes a scenario in which an aircraft de-trail and other is landing. De-trail is the event that occurs when an aircraft loss its assigned track (heading angle) that may be caused due to mechanical failure, external disturbances, aircraft upset conditions, or inappropriate crew actions or responses which is recommended to be resolved by technique to assist de-trailing aircraft on current runway and assign an alternate runway to other aircraft.

## 5.1. Hierarchical Petri Nets Model of ATC Metaverse

Since Metaverse is an integration of several distributed concurrent units, the formal model of ATC Metaverse is developed by using hierarchical CPN, which have been specified in chapter 4. The ATC model, as discussed in the previous section, is transformed into a hierarchical Petri net model through the application of powerful features offered by hierarchical CPN.

### 5.1.1. Interconnection of ATC Metaverse Components

In Figure 5 main hierarchical CPN model of ATC metaverse show interconnection between six substitution transitions represented by blue rectangles. In Figure 5, rectangles represent substitution transitions, which allow the construction of hierarchical reconfigurable Petri nets by breaking down complex nets into simpler, modular components, thereby providing a detailed, multi-layered model of the system. Creating large, intricate nets can be a cumbersome task. But similar to modular programming, the construction of CPNs can be broken into smaller pieces by utilizing the facilities within CPN Tools for creating substitution transitions. A hierarchical reconfigurable Petri net uses substitution transitions to implement the hierarchy. A substitution transition is a special kind of transition that itself does not fire, instead it contains a subnet that defines the behaviour that takes place in its stead. Conceptually, nets with substitution transitions are nets with multiple layers of detail can have a somewhat simplified net that gives a broad overview of the system's model, and by substituting transitions of this top-level net with more detailed pages set as subpages for substitution transitions can bring more and more detail into the model.

There are six places that are represented by ellipses, and the tokens in initial places are represented by circles. A place can contain any number of tokens, depicted as green circles in Figure 5. In a standard Petri net, tokens are indistinguishable whereas in CPN, every token has a value represented by green rectangles as shown Figure 5. For example, 1`("ATC","DATA",0) represents there is one token of type PACKET in which first value for sender is set as "ATC", i.e., it is send by main CPN, second value for data is set as "DATA", i.e., it contains arbitrary value. The third value for command type is set to 0 i.e., "Create Virtual Environment". It is important to note that, as the token fires through the subnets the value of sender changes respectively, but data and command type will remain the same. The diagrams shown in Figure 5 is a CPN model developed in CPN Tools which is a widely used tool in which values of tokens are typed, and can be tested (using guard expressions) and manipulated with a functional programming language. Tokens represent system's state and move through places and transitions to model dynamic behaviour and transitions fire based on token availability, depicting flow of information or resources within system.

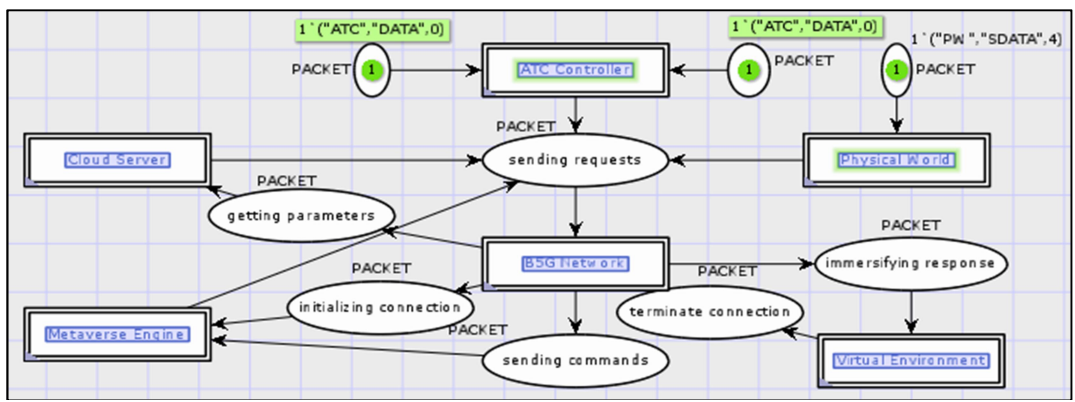

**Fig. 5.** Main hierarchical Petri net developed for the ATC Metaverse.

*ATC controller* initializes connection by sending a request to *Metaverse Engine* through *B5G Network*. After getting authentication from *Metaverse Engine* and sensor inputs from *Physical World*, virtualization is created by *Virtual Environment*. *ATC controller* can send a command to *Metaverse Engine* through *B5G Network*, which processes the command and send the request back for immersification to *Virtual Environment* via *B5G Network*. In case of any conflicting situation as defined in Table 4, *ATC Controller* can send a command to get recommendations. In this scenario, *Cloud Server* gets parameters from *Metaverse Engine* through *B5G Network*. It is assumed that all other subsystems set their states contingent upon the controller's authorization, which is pivotal in ensuring coordinated and conflict-free operation. Similarly, when *ATC Controller* wishes to exit the Metaverse, he sends a command to *Virtual Environment* to terminate the connection.

### 5.1.2. ATC Controller

This subnet formalizes the ATC controller (Figure 6). A controller can initiate a request to connect to the Metaverse, which is then sent to the Metaverse engine through the output port. Similarly, the controller can also issue two concurrent commands to the Metaverse engine, one for the movement of the avatar and one command initiated through the controller's haptic. It is important to note that the third attribute of a PACKET is defined to determine the type of command (i.e., 0 for create virtual environment, 1 for avatar movement, 2 for getting recommendations, and 3 for getting cached tutorials). The important and helpful tutorials essential for guidance of ATC controllers are cached in the Metaverse Engine for quick retrieval. In the Fig-6, the s, d and t are the variables of type STRING, STRING and INT respectively. These variables are declared to receive the values of PACKET type token from an input port where s holds first value that is the receiver abbreviation, d holds second value which contains the data and t holds third value that is a command denoted by an integer value. The values of these variables are transmitted throughout the subnet via arcs to the places/transitions for those who require them for processing.

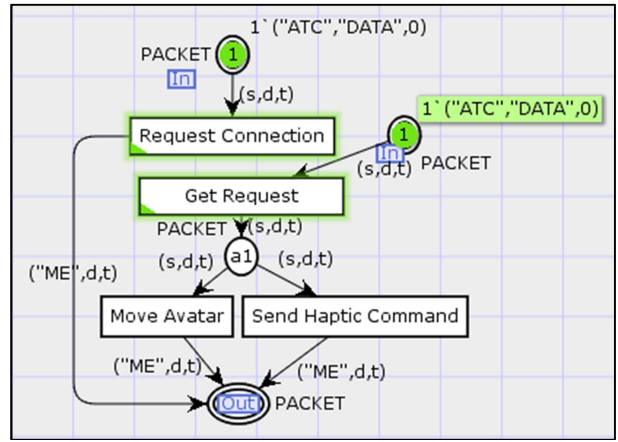

**Fig. 6.** Subnet ATC controller developed for the ATC Metaverse.

### 5.1.3. B5G Network

The ultra-high-speed Internet called B5G (Beyond 5G also called 6G) is abstracted through a subnet named B5G Network (Figure 7). The network receives a packet from the sender and transmits it to its receiver. This subnet receives command via input port **In** from ATC controller. After receiving signals, the data is retrieved and signals are transmitted towards the recipient subnet via the output ports. The output ports are defined as, when there is a command is to initialize avatar, a packet is sent to subnet Metaverse Engine. Similarly, output ports for other commands such as move avatar, get recommendations etc., a packet is sent to subnet Metaverse Engine and a packet is sent to subnet Virtual Environment respectively.

In the Figure 7, the variables *s, d* and *t* are of type STRING, STRING and INT respectively. These variables are declared to receive the values of PACKET type token from an input port where *s* holds first value that is the receiver abbreviation, the variable *d* holds second value which contains the data, and the variable *t* holds third value that is a command denoted by an integer value. The values of these variables are transmitted throughout the subnet via arcs to the places/transitions for those which require them for processing.

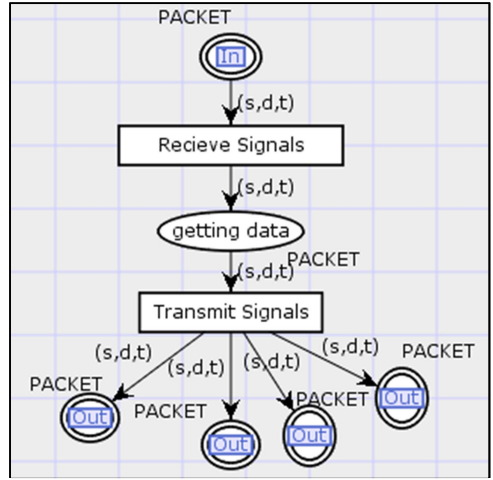

**Fig. 7.** Subnet B5G Network developed for the ATC Metaverse.

#### 5.1.4. Metaverse Engine

It is the hub of the Metaverse system that is formalized by the subnet named Metaverse Engine (Figure 8). For a given input sent by an ATC controller through the B5G network via a packet, three types of commands can be executed at a time. Process motion is executed to move the avatar and digital twin of the aircraft controlled. It is done locally by using the system's resources through the sensor's input from the physical world. ATC controller may send a request to get safety recommendations from the Metaverse engine, which requires extensive computations. In turn, it needs to interact with the cloud server for its processing. The controller can also request tutorials that are cached on the local system. In Fig 8, there are two incoming tokens received via input ports. One token **In** from subnet **ATC controller** for receiving the command to be executed, and another token **In** from subnet **Physical Environment** for getting sensors data.

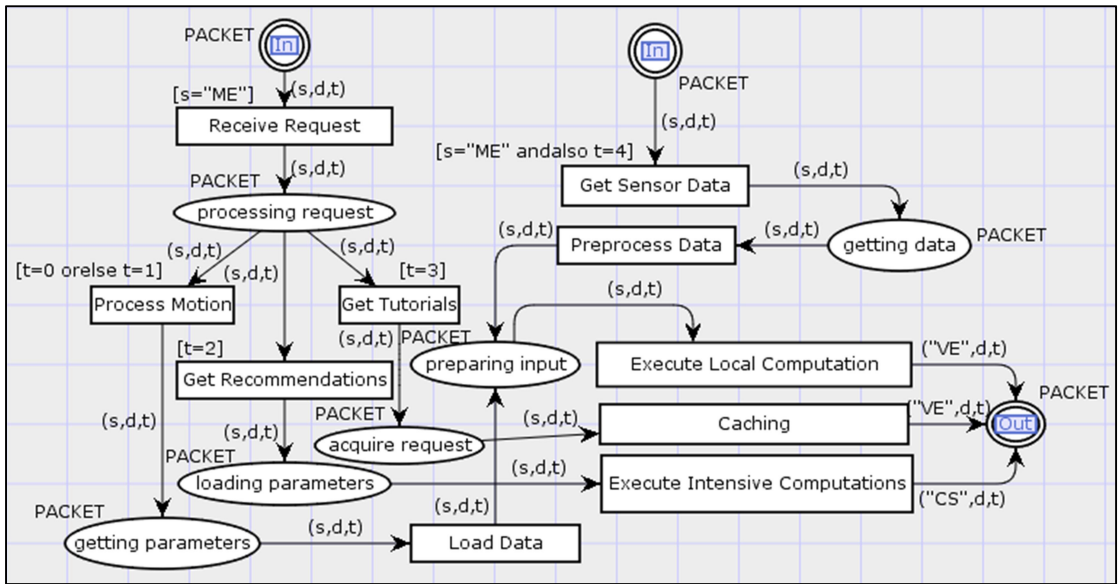

**Fig. 8.** Subnet Metaverse Engine developed for the ATC Metaverse.

#### 5.1.5. Virtual Environment

The virtual environment is represented by a subnet named Virtual Environment (Figure 9). After authentication by the Metaverse engine, the virtual environment creates a digital ecosystem through the creation of the ATC controller's avatar and digital twins of all aircraft being controlled by him. Likewise, based on the response received from the Metaverse engine, the virtual environment is refreshed till the controller ends the session.

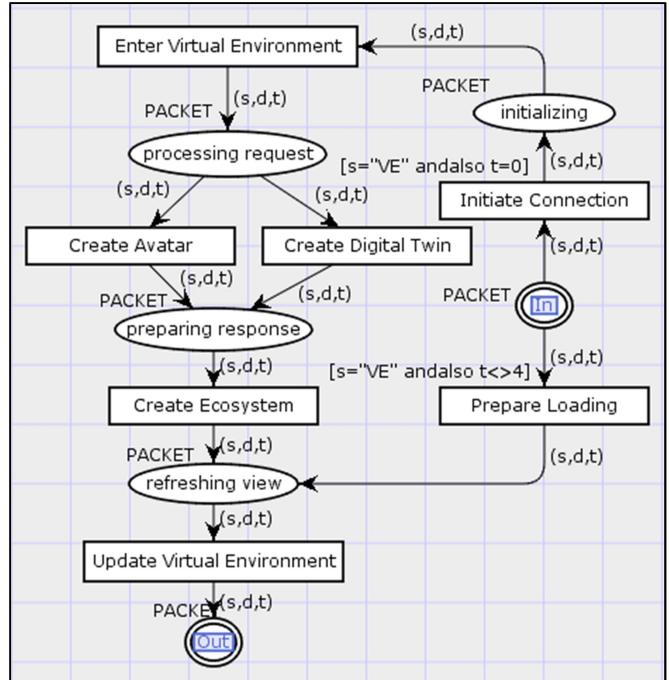

**Fig. 9.** Subnet Virtual Environment developed for the ATC Metaverse.

### 5.1.6. Physical Environment

The formalization of the physical world is shown by the subnet Physical World (Figure 10). The physical world is a real-world sensor network, and the sensor data is received from sensors and transmitted to the network.

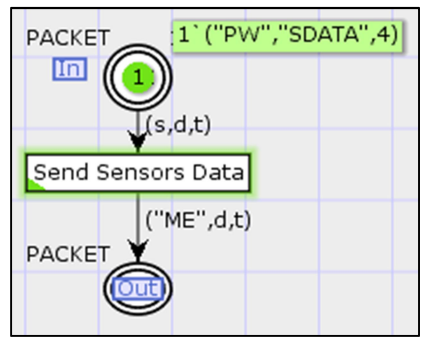

**Fig. 10.** Subnet Physical World developed for the ATC Metaverse.

### 5.1.7. Cloud Server

The formalization of the Cloud Server is shown by the subnet Cloud Server (Figure 11). Based on the input from the Metaverse engine, the cloud offloads the task of computation, which makes predictions and generates results that are sent to a virtual environment for display. This subnet is invoked when Metaverse engine seeks an advice in case of a conflicting situation which is ensured by defining a guard condition [s="CS"] on transition Cloud Offloading. In order to predict a suitable recommendation for resolving conflict the cloud server offloads this task for necessary computation and predictions. Based on its output, results are processed to generate conflict resolution suggestions which are sent to virtual environment for display.

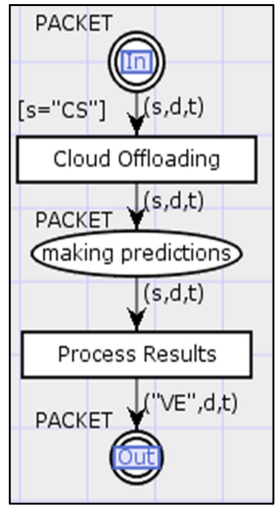

**Fig. 11.** Subnet Cloud Server developed for the ATC Metaverse.

### 5.2. Verification of ATC Metaverse Model

CPN Tools are a toolset for analysing the dynamic behaviour of a system's model through a deep inspection of critical properties such as concurrency, synchronization, and conditional flows, through a token-passing mechanism [58]. The toolset also supports extensibility, as it allows a modeler, to write their own theorems and properties in a Meta Language (ML) and to prove these theorems by using the tool's built-in theorem provers. Therefore, CPN Tools are selected for verification of the ATC Metaverse. The formal ATC Metaverse is analysed through a state space graph, and, then, certain properties like reachability, liveness, and boundedness are explored. Further, theorems are described and proved for exploring important properties of concurrency, safety, and liveness.

#### 5.2.1. ATC Metaverse in State Space Graph

A state space graph is a directed graph, in which nodes represent a set of reachable markings, and edges are represented by occurring binding elements. The formal ATC Metaverse generated a state space consisting of 210 nodes and 553 arcs, while Figure 12 only shows the first few nodes and their interconnections. It is proved by the toolset CPN Tools that the reachability of each state occurs effectively, although there are many interacting states from different subnets. No unreachable state has been identified.

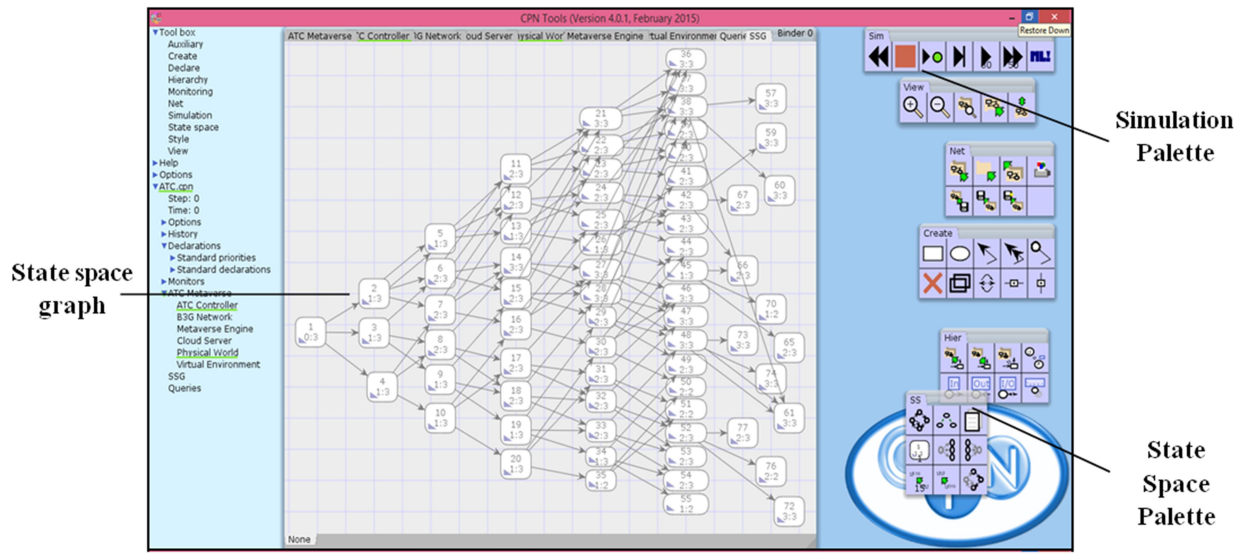

**Fig. 12.** State space graph of the ATC Metaverse model.

### 5.2.2. Liveness Analysis of ATC Metaverse

Liveness is the property that asserts a desirable state is eventually reached by a system during its state space. Liveness is used to verify that all those states stay alive, which is important for the execution of a particular property. In this research, an extensive liveness analysis is performed for verification of the formal ATC Metaverse model, including dead marking, dead transitions, and live transitions. A marking is dead, if it has no enabled transitions; a transition is termed as dead instance, if it is disabled in every reachable marking; and, a transition is termed as live instance, if it is enabled in some reachable marking. The formal ATC model is executed for every input. Table 5 shows the results of liveness analysis of the ATC Metaverse model, in which the first column shows values given as input to the Metaverse engine subnet layer (as shown in Figure 4). The descriptions of the input values are narrated in the second column. The third column expresses the details of the resulting state graph (i.e., nodes and arcs), and column four presents the identified dead states in the corresponding subnets. For example, when input 2 "Get recommendations using deep neural network" is sent by the ATC controller subnet, then certain states of the Metaverse engine (e.g., Caching, Get Tutorials) and certain states of the Virtual Environment (e.g., Create Avatar, Initiate Connection) are identified as dead states. Those states must be dead logically, as they have not been involved in the given input. Similarly, for all other inputs given to the formal ATC Metaverse model, as shown in Table 4, only those dead states, which must be dead during the execution of a particular input, are identified. Therefore, the proposed ATC Metaverse model demonstrates correct behaviour from the perspective of liveness.

**Table 5.** Results of liveness analysis of the ATC Metaverse.

| Input | Description | State Space Graph | | Dead States | | |
| --- | --- | --- | --- | --- | --- | --- |
| | | Nodes | Arcs | Cloud Server | Metaverse Engine | Virtual Environment |
| 0 | Create Virtual Environment | 2235 | 7571 | Cloud Offloading | Caching | nil |
| | | | | Process Results | Execute Intensive Computations | |
| | | | | | Get Recommendations | |
| | | | | | Get Tutorials | |
| 1 | Update Virtual Environment | 1515 | 4754 | Cloud Offloading | Caching | Create Avatar |
| | | | | Process Results | Execute Intensive Computations | Create Digital Twin |
| | | | | | Get Recommendations | Enter Virtual Environment |
| | | | | | Get Tutorials | Initiate Connection |
| | | | | | | Create Ecosystem |
| 2 | Get recommendations using deep neural network | 2235 | 7061 | nil | Caching | Create Avatar |
| | | | | | Load Data | Create Digital Twin |
| | | | | | Get Tutorials | Create Ecosystem |
| | | | | | Process Motion | Enter Virtual Environment |
| | | | | | | Initiate Connection |
| 3 | Get tutorials | 1305 | 4083 | Cloud Offloading | Execute Intensive Computations | nil |
| | | | | Process Results | Get Recommendations | |
| | | | | | Load Data | |
| | | | | | Process Motion | |

### 5.2.3. Boundedness Analysis of ATC Metaverse

The boundedness property states the number and type of tokens a place can hold in all reachable markings of the net. Boundedness is applied to verify that the model remains bound for every possible initial marking throughout the entire state space. We have performed two types of boundedness analysis for the verification of formal ATC Metaverse: integer bound and multi-set bound. An integer bound (I) is the number of tokens in a place, while ignoring the token colours, whereas a multi-set bound (M) is the number of tokens in a place by providing information about possible token colours. The ATC Metaverse model is executed for every input, and the boundedness values of each subnet are identified. Table 6 shows the results of the boundedness analysis of the ATC Metaverse, in which the first column shows all subnets, whereas the remaining columns show the boundedness

23values at given inputs for integer and multi-set bounds. It has been noticed that the unreachable subnets for a specific input remained unbounded with a value 0 for both integers bound and multi-set bound. For example, when input 3 "Get Tutorial" is sent by the ATC controller subnet, the Cloud Server subnet remains 0 for both integers bound and multi-set bound. It shows that the Cloud Server subnet is not reachable for this particular input, and it must be unbound, as this subnet is not involved in the execution of the given input. Similarly, all other inputs given to the ATC Metaverse model (Table 6) identified only those states unbounded, which must be reachable during the execution of a particular input. Therefore, the proposed ATC formal model demonstrated correct behaviour from the perspective of boundedness.

Table 6. Results of boundedness analysis of formal ATC Metaverse

| Input | 0 | | 1 | | 2 | | 3 | |
|---|---|---|---|---|---|---|---|---|
| Subnets | I | M | I | M | I | M | I | M |
| ATC Metaverse | 9 | 9 | 9 | 9 | 9 | 9 | 1 | 1 |
| ATC Controller | 1 | 1 | 1 | 1 | 1 | 1 | 9 | 9 |
| Virtual Environment | 4 | 4 | 4 | 4 | 4 | 4 | 4 | 4 |
| Metaverse Engine | 6 | 6 | 6 | 6 | 6 | 6 | 4 | 4 |
| Cloud Server | 1 | 1 | 1 | 1 | 1 | 1 | 0 | 0 |
| B5G Network | 1 | 1 | 1 | 1 | 1 | 1 | 1 | 1 |
| Physical World | 0 | 0 | 0 | 0 | 0 | 0 | 4 | 4 |

#### 5.2.4. Model Checking

Some important system critical properties of a model such as concurrency, safety, and liveness of the ATC Metaverse have mathematically been defined and proved. For this, a model checker, which exhaustively explores all possible states of a finite-state system model, checks that the model meets a given property. The process of model checking ensures that a formal specification meets the stated properties. CPN Tools offers a powerful way to check system properties by using standard Meta language (ML) as an input language. ML is executed by an ASK-CTL model checker, to ensure the overall model correctness [40], including concurrency, safety, and reachability.

*a.* **Concurrency Property**

Concurrency is the property of a model, in which two or more control flows are executed simultaneously without following any order. We have mathematically defined a concurrency property about the ATC Metaverse, which states that *"Initialization and refreshing view of the virtual environment cannot occur concurrently"* (Figure 13). It means that in the subnet Virtual Environment, initialization and refresh cannot occur at the same time.

```
fun NonConcurrentMarking n = ((Mark.Virtual_Environment'initializing 1 n) == 1`("VE", "Data", 0)) andalso
                              ((Mark.Virtual_Environment'refreshing_view 1 n)==1`("VE", "Data", 1));
val ConcurrencyProp = INV(NOT(NF("INITIALIZATION AND REFRSHING VIEW CANNOT OCCUR CONCURRENTLY",
                       NonConcurrentMarking)));
eval_node ConcurrencyProp InitNode;
```

**Fig. 13.** Formalization of the concurrency property in the ATC Metaverse model.

The defined concurrency property is evaluated via Execute ML command of state space tool, which investigates whether the state space contains independent paths with non-overlapping states. First, it is evaluated from a place named initializing of subnet Virtual Environment when token contains a command to initiate connection, and another path from a place named refresh view of subnet Virtual Environment when token contains a command to move avatar. If the result of the evaluation process returns true, it will prove that the defined concurrency property holds for the entire state space of the developed ATC model and there exists two aforementioned concurrent paths with non-overlapping states.

*b.* **Safety Property**

Safety is the property that is concerned that the model never enters into any harmful state during its entire state space, ensuring that nothing bad happens during execution. We have mathematically defined a safety property about the formal ATC Metaverse, which states that *"A Cloud Server makes predictions, if and only if Metaverse Engine sends all required parameters"* (Figure 14). It means that in case of any conflicting situation, the subnet Cloud Server cannot make its predictions until the subnet Metaverse Engine loads all required parameters.

```
fun BeginMarking n = ((Mark.Metaverse_Engine'loading_parameters 1 n) == 1`("ME", "Data", 2));
fun EndMarking n = ((Mark.Cloud_Server'making_predictions 1 n) == 1`("CS", "Data", 2));
val SafetyProp = POS(AND(NF("Metaverse Engine receives get recommendations command ", BeginMarking),
            EXIST_UNTIL((NF("",BeginMarking)),
                  NF("cloud server makes predictions", EndMarking))));
```

**Fig. 14.** Formalization of safety property for the ATC Metaverse model.

This property is evaluated via Execute ML command of state space tool which investigates whether, the state space contains a directed path from a place named loading parameters of subnet Metaverse Engine. It is evaluated, when token contains a command for getting recommendations to a place named make predictions of subnet Cloud Server, and when token contains a command for getting recommendations. If the result of evaluation process returns true, it proves that there exists an occurrence sequence from source to destination marking, and the developed safety property holds for the entire state space of the developed ATC model.

*c.* **Reachability Property**

Liveness is the property that is concerned that a model eventually enters into any intended state during its entire state space, ensuring that something good happens during execution. We have mathematically defined a liveness property about the formal ATC Metaverse, which states that *"Virtual environment is closed when ATC controller ends session"* (Figure 15). It means that the subnet Virtual Environment is immediately terminated, when the subnet ATC controller closes its session.

```
fun RequiredMarking n = ((Mark.Virtual_Environment'closed 1 n) == 1`("ME", "Data", 5));
val ReachabilityProp = EV(NF("Virtual Environment is closed when controller ends session ",
                RequiredMarking));
eval_node ReachabilityProp InitNode;
```

**Fig.15.** Formalization of the reachability property for the ATC Metaverse model.

This property is evaluated via Execute ML command of state space tool which investigates, whether state space contains a directed path from the initial node to a place named closed of subnet Virtual Environment when token contains a command for connection termination. The result of evaluation process returns true, it proves that there exists an occurrence sequence from source to destination marking, and the defined reachability property holds for entire state space of the developed ATC model.

## 6. Overall Discussion

In this research work, a framework is presented that utilizes formal modelling techniques to model complex applications of Metaverse. The case study of Air Traffic Control system is modelled, in order to demonstrate the usability of the presented framework. This research provides a starting point for formal modelling and verification Metaverse applications. It will serve as a template for modelling and describing various important Metaverse properties that must be formally verified. Furthermore, the proposed framework will be useful to increase the reusability of the Metaverse's formal specifications and verifications during the modelling of multiple Metaverse applications and will also help to address the problem of increased complexity of Metaverse applications. The Metaverse applications exhibit certain properties (such as parallelism, synchronization, distribution, data sharing, and non-determinism), which are difficult to model by using conventional formal modelling techniques (e.g., differential equations and difference equations).

The application of Petri-nets in the formal modelling of the Metaverse applications will help to effectively describe the aforementioned inherent properties through robust and powerful mathematical notations. As compared to other graphical formal modelling tools, like block diagrams or logical trees, the use of hierarchical CPN allows to model the coordinating units of a Metaverse in the form of subnets, which provided means for efficient model analysis. CNP combines Petri Nets with a programming language that allow us to obtain a scalable modelling language for concurrent systems. The use of CPN provides a foundation for modelling concurrency and synchronization. The associated programming language provide primitives for modelling, data manipulation and creating compact and parameterizable models. Further, during the modelling of ATC metaverse application, CPN allows to create a hierarchical net which break its complexity, by dividing it into a number of sub-models. The use of CPN allows us to organize the complex metaverse application in simple and understandable way.

In addition, this research provides the formal modelling of a complex safety-critical ATC Metaverse, which is formally verified and discussed in detail. Furthermore, from the modelling of ATC Metaverse, it can be concluded that the developed Metaverse framework can be consistently applied to every Metaverse application. For instance, the proposed metaverse framework is suitable for centralized blockchain applications as well in which metaverses allow interactions to evolve within virtual spaces. This creates a complete digital economic environment in which users monetize their acquisitions and creations. In addition, the proposed metaverse architecture can be utilized for DAO (Decentralized Autonomous Organization) in which smart contracts will build the rules within a given Metaverse. In decentralized blockchain Metaverses application, the decision-making power does not lie with a central organization, but with the users. Each user who owns a token of the Metaverse carries out a decisional role on the administration of the virtual world in which he evolves.

The applicability of the proposed Metaverse framework should be evaluated by modelling other complex Metaverse applications. Since the Metaverse is a complex and decentralized integration of different components, the application of formal modelling in this research helps to ensure a deep system insight into system requirements. Similarly, the process of formal verification helps us to define and prove some critical system properties, like state space, liveness, and, boundedness, which provide guarantees and assurances about the correctness of Metaverse applications. For instance, liveness ensures that the air traffic control system remains fully operational, allowing all aircraft to eventually receive necessary clearances and preventing indefinite waiting or deadlock situations. Meanwhile, reachability ensures that the system can achieve any necessary state from any given state, enabling controllers to manage dynamic air traffic scenarios safely and efficiently. These properties are then formally verified, which in itself is a challenging task. Formal verification of CPN model is a process of checking whether a design satisfies required properties. It helps to confirm that our system models behave correctly through application of mathematically rigorous procedures and search through possible execution paths of our model to ensure system correctness. In addition, the use of a powerful model checker offered for Petri-Nets helps to explore properties of our developed Metaverse through verification of all interleaving and parallel branches, which was not possible by conventional simulators applied during the testing phase of the model.

## 7. Conclusion

### 7.1. Summary and Conclusion

This paper serves as a baseline for the integration of formal methods into Metaverse applications, specifically, the most significant contribution of this work is the development of a framework, which can be applicable to diverse Metaverse applications. The second important contribution of this research work, is the definition of an abstract ATC Metaverse application, which is then transformed into a corresponding formal model by using hierarchical CPN. Finally, the proposed Petri-net-based model is verified and analyzed through methods offered by CPN Tools.

### 7.2. Limitations and Future Directions

One of the limitations of this research work is that, a component named 'Security and Privacy' is presented in the proposed Metaverse architecture, but not applied and integrated into a Metaverse application. It has been left for future work. Regarding the hierarchal CPN, one of the promising drawbacks observed during the using CPN for metaverse applications is that it faces state explosion problem. For highly complex metaverse applications, it is anticipated that the state space will become very large and it may not be fully constructed.

Another future work is related to the use of Petri nets for the formalization of the Metaverse components and Metaverse applications. In future work, it is worth to be investigated how Petri nets can be linked with other formal methods, like axiom-based, structural-based, and logic-based formal methods. Similarly, to handle the complexity of interactions the dynamic modelling techniques, hybrid approaches, scalability, real-time data integration, and regular validation can be considered to better capture the dynamic nature of the Metaverse. With that, it might be possible to identify subtle errors during the early stages of application developments.

In continuation of this study, it is also planned to define some more critical properties (besides concurrency, safety, and reachability) for the hierarchical CPN. It will enhance the confidence in terms of correctness of a system to be developed. Furthermore, the strengths of Petri nets may be further evaluated by modelling more complex applications in the Metaverse from the real world by modelling of other Metaverse applications ranging from complex safety-critical systems to simple gaming applications.

26## References

1. A. Coronato, and De Pietro, G. (2011). Formal specification and verification of ubiquitous and pervasive systems. ACM Transactions on Autonomous and Adaptive Systems (TAAS), 6(1), 1-6.

2. A. Hall (2007). Realising the Benefits of Formal Methods. J. Univers. Comput. Sci., 13(5), 669-678.

3. A. Houser, Ma, L. M., Feigh, K. M., and Bolton, M. L. (2018). Using formal methods to reason about taskload and resource conflicts in simulated air traffic scenarios. Innovations in Systems and Software Engineering, 14, 1-14.

4. A. Jarrar and Balouki, Y. (2018). Towards sophisticated air traffic control system using formal methods. Modelling and Simulation in Engineering, 2018.

5. A. Jarrar, Ait Wakrime, A., and Balouki, Y. (2020). Formal approach to model complex adaptive computing systems. Complex Adaptive Systems Modeling, 8(1), 3.

6. A. Martín Montes, Burbano Cendales, A. M., and León de Mora, C. (2017). An Intelligent Methodology for Modeling Semantic Knowledge in Industrial Networks. WSEAS Transactions on Computers, 16, 179-188.

7. A. Rehman (2021). Machine learning based air traffic control strategy. International Journal of Machine Learning and Cybernetics, 12, 2151-2161.

8. A. Siyaev and Jo, G. S. (2021). Towards aircraft maintenance metaverse using speech interactions with virtual objects in mixed reality. Sensors, 21(6), 2066.

9. A. Souri, Rahmani, A. M., Navimipour, N. J., and Rezaei, R. (2020). A hybrid formal verification approach for QoS-aware multi-cloud service composition. Cluster Computing, 23, 2453-2470.

10. B. Kye, Han, N., Kim, E., Park, Y., and Jo, S. (2021). Educational applications of metaverse: possibilities and limitations. Journal of educational evaluation for health professions, 18.

11. B. Vogel-Heuser, Huber, C., Cha, S., and Beckert, B. (2021, July). Integration of a formal specification approach into CPPS engineering workflow for machinery validation. In 2021 IEEE 19th International Conference on Industrial Informatics (INDIN) (pp. 1-8).

12. C. B. Fernandez and Hui, P., 2022, July. Life, the Metaverse and everything: An overview of privacy, ethics, and governance in Metaverse. In 2022 IEEE 42nd International Conference on Distributed Computing Systems Workshops (ICDCSW) (pp. 272-277). IEEE.

13. C. Mahmoudi, Mourlin, F., and Battou, A. (2018, April). Formal definition of edge computing: An emphasis on mobile cloud and IoT composition. In 2018 Third international conference on fog and mobile edge computing (FMEC) (pp. 34-42).

14. D. Ivanov and Dolgui, A., 2021. A digital supply chain twin for managing the disruption risks and resilience in the era of Industry 4.0. Production Planning and Control, 32(9), pp.775-788.

15. D. L. Parnas (2010). Really rethinking' formal methods'. Computer, 43(1), 28-34.

16. D. Liu, Zhu, H., Xu, C., Bayley, I., Lightfoot, D., Green, M., and Marshall, P. (2016, June). Cide: An integrated development environment for microservices. In 2016 IEEE International Conference on Services Computing (SCC) (pp. 808-812).

17. F. Li, 2020. Leading digital transformation: three emerging approaches for managing the transition. International Journal of Operations and Production Management, 40(6), pp.809-817.

18. F. Li, 2020. The digital transformation of business models in the creative industries: A holistic framework and emerging trends. Technovation, 92, p.102012.

19. G. Vial, 2021. Understanding digital transformation: A review and a research agenda. Managing Digital Transformation, pp.13-66.

20. H. Garavel, Beek, M. H. T., and Pol, J. V. D. (2020). The 2020 expert survey on formal methods. In Formal Methods for Industrial Critical Systems: 25th International Conference, FMICS 2020, Vienna, Austria, September 2–3, 2020, Proceedings 25 (pp. 3-69). Springer International Publishing.

21. H. Ning, Wang, H., Lin, Y., Wang, W., Dhelim, S., Farha, F., ... and Daneshmand, M. (2021). A Survey on Metaverse: The State-of-the-art, Technologies, Applications, and Challenges. arXiv preprint arXiv:2111.09673.

22. https://formalmethods.fandom.com/wiki/Companies (last visited: 06-02-2023)

23. https://github.com/Maryam-IIUI/FM-Metaverse.git

24. https://www.frequentis.com/en/air-traffic-management/digital-experience

25. https://www.reuters.com/technology/boeing-wants-build-its-next-airplane-metaverse-2021-12-17/

26. I. A. Akour, Al-Maroof, R. S., Alfaisal, R., and Salloum, S. A. (2022). A conceptual framework for determining metaverse adoption in higher institutions of gulf area: An empirical study using hybrid SEM-ANN approach. Computers and Education: Artificial Intelligence, 3, 100052.

27. I. Cafezeiro, Viterbo, J., Rademaker, A., Haeusler, E. H., and Endler, M. (2014). Specifying ubiquitous systems through the algebra of contextualized ontologies. The Knowledge Engineering Review, 29(2), 171-185.